\documentclass[12pt]{article}
\usepackage[T1]{fontenc}
\usepackage[utf8]{inputenc}
\usepackage{amssymb,amsmath}
\usepackage{natbib}
\usepackage{graphicx}
\usepackage[unicode=true]{hyperref}

\usepackage{longtable}
\usepackage{booktabs}
\usepackage{CJKutf8}  
\usepackage{pifont}  

\hypersetup{colorlinks=true, citecolor=blue, linkcolor=blue, urlcolor=blue, filecolor=black} 

\usepackage[onehalfspacing]{setspace}
\usepackage{fullpage}
\usepackage{times}
\usepackage{microtype}

\title{Public Domain Rank: Identifying Notable Individuals with the Wisdom of the Crowd}
\author{Allen B. Riddell\\ {\small Neukom Institute, Dartmouth College}}
\date{\footnotesize \today \\ Version 1.0}

\begin{document}

\maketitle

\newcommand{\fix}{\marginpar{FIX}}
\newcommand{\new}{\marginpar{NEW}}
\newcommand{\logit}{\text{logit}}
\newcommand{\numwikipediaindividuals}{1,011,304}
\newcommand{\numwikipediaindividualstrain}{85,424\ }
\newcommand{\numgutenberg}{45,000\ }
\newcommand{\numtopics}{200\ }
\newcommand{\numtranslations}{10\ }
\newcommand{\numfeatures}{233\ }  
\newcommand{\percenthavingdigitaleditions}{3\%}
\newcommand{\websiteurl}{http://publicdomainrank.org}
\newcommand{\woolfrank}{1,211}
\newcommand{\woolfscore}{94}
\newcommand{\amisanirank}{565,205}
\newcommand{\amisaniscore}{48}
\newcommand{\medianmodernlibrary}{4,107\ }

\begin{abstract}

Identifying literary, scientific, and technical works of enduring interest is
challenging. Few are able to name significant works across more than
a handful of domains or languages.  This paper introduces an automatic method
for identifying authors of notable works throughout history. Notability is
defined using the record of which works volunteers have made available in public
domain digital editions. A significant benefit of this bottom-up approach is that
it also provides a novel and reproducible index of notability
for all individuals with Wikipedia pages. This method promises to supplement the
work of cultural organizations and institutions seeking to publicize the
availability of notable works and prioritize works for preservation and
digitization.

\end{abstract}

\begin{CJK}{UTF8}{min}
\section{Introduction}\label{introduction}

Every year thousands of literary, scientific, and technical works enter the
public domain. Works in the public domain are unencumbered by restrictions that
hinder their being used, shared, and re-purposed (``remixed'').
A high school teacher in Australia may assign students George
Orwell's \emph{Nineteen Eighty-Four} knowing that students will be able to
obtain a copy online without cost. A theater company in Canada may stage a play
by Albert Camus without requesting---and potentially being refused---permission
from the writer's estate. Evidence of interest in public domain works is not
difficult to find. Recent years have witnessed the proliferation of organizations
committed to preserving and making accessible works in the public domain. Prominent examples
of such organizations include the Internet Archive,
\href{http://gutenberg.org}{Project Gutenberg},  and
\href{http://librivox.org/}{Librivox}.\footnote{Project Gutenberg has over
\numgutenberg works in its collection. Project Gutenberg Canada and Distributed
Proofreaders Canada have produced over 1,000 works.  Librivox, established in
2005, now has over 6,244 free audiobooks recorded by volunteers, principally of
public domain books. The Internet Archive, which hosts a variety of public domain
materials, is ranked among the top 200 websites in the world. Project Gutenberg
ranks in the top 10,000. Millions visit these sites every month. (Rankings
collected from Alexa on October 4, 2013.)}

Every year those interested in publicizing the availability of public domain
materials face the challenge of identifying notable works that will enter the public
domain in the coming year.  Notable works are those in which there is an enduring interest.  Such
works have ``stood the test of time'' and command, decades after their
publication, a significant contemporary following.  (Notable individuals are
analogously defined.) Identifying such works is difficult because interest in
works and their authors depends on subject matter, geography, and language. For
example, the community of readers interested in the works of the American author
Flannery O'Connor (1925-1964) and the community interested in the works of the
Chinese writer Lao She (老舍) (1899-1966) are not identical. For this reason,
assembling even a rudimentary list of notable works entering the public domain
requires a capacious knowledge of culture and science. Existing efforts rely
on volunteers to trawl through lists of authors
in search of notable works.\footnote{Personal communication
with Communia Association members Alek Tarkowski and Primavera De Filippi, March
21, 2013.} In 2011, authors identified by the Communia Association's ``Public
Domain Day'' included Walter Benjamin, Isaac Babel, F. Scott Fitzgerald, and
Emma Goldman \citep{communia_association2011public}.  (The Association's
``Public Domain Day'' identifies authors whose works are entering the public
domain in 70 year \emph{post mortem auctoris} (``life plus 70 years'') copyright
regimes.) Another significant
``Public Domain Day'' is organized by the US-based
\href{http://web.law.duke.edu/cspd/publicdomainday}{Center for the Study of
the Public Domain} \citep{jenkins2013ambiguous}.\footnote{As
    \citet{jenkins2013ambiguous} notes, ``Public Domain Day'' in the United
States is an exercise in counter-factual thinking as no works will enter the
public domain in the US until 2019 at the earliest.} The authors and works
featured in the Public Domain Day collections tend to be those of interest to
audiences geographically or linguistically connected to the sponsoring organization.
While few would object to the lists of notable authors
identified by these organizations, the selection procedures are typically
opaque and depend on the judgements of the handful of individuals involved.

Identifying notable works that have been in the public domain for decades is, by
comparison, straightforward. We have the empirical record of what works
volunteers have edited and published in online repositories such as Project
Gutenberg. In the deliberations of these volunteers, we have a valuable
independent judgement of which works (and, by extension, which authors) have
stood the test of time.  Unfortunately, this judgement is only reliable for
works that have been in the public domain for a considerable amount of time; the
collective judgement of the crowd is unavailable for works still covered by
copyright monopolies.

This paper introduces and evaluates an automatic method for approximating this
collective judgement when it is unavailable. Using data from Wikipedia and the Online Books Page,
individuals are ranked in terms of how strongly they resemble individuals whose
works have been published in freely available digital editions.

There are two major applications of this ranking. First, the \emph{Public Domain Rank}
promises to supplement the labors of organizations and libraries seeking to
publicize the availability of notable works in the public domain and to
prioritize works for preservation and digitization. A second application
arises from treating the Public Domain Rank as a general, independent index of
an individual's importance for contemporary audiences. For example, Wikipedia
editors stand to benefit from being able to identify ``overlooked''
individuals---those whose biographical articles do not adequately reflect their
importance to existing communities.

\nocite{lessig2008remix}

\section{Curating the Public Domain: Project Gutenberg and The Online Books Page}\label{data}

For the subset of published works that have been in the public domain for many
years, we benefit from an unambiguous signal of a work's importance for
contemporary audiences: the existence of a freely available \emph{digital
edition}. The prototypical digital edition is a Project Gutenberg edition of
a work. Digital editions involve considerable human labor beyond page scanning,
such as manual entry and proofreading.\footnote{Texts derived from manually
corrected OCR are counted as digital editions.} Because the creation of
a digital edition is typically volunteer-driven, time-consuming, and
labor-intensive, the existence of a digital edition of a work is a strong signal
that a work commands a contemporary following.

With over \numgutenberg texts, Project Gutenberg figures among the most
significant repositories of public domain digital editions. Many other
collections exist, such as
\href{https://ebooks.adelaide.edu.au/}{eBooks@Adelaide} and
\href{http://www.gutenberg.ca/}{Project Gutenberg Canada}. Works in these and other
collections are assembled in a meta-index,
\href{http://onlinebooks.library.upenn.edu}{The Online Books Page}.
Hosted at the University of Pennsylvania Libraries and curated by John Mark
Ockerbloom, the Online Books Page draws on a broad range of sources for its index of over
a million books.\footnote{The Online Books Page includes references to books available online
that are, by the definition used in this paper, not digital editions.  For
example, books for which only page scans are available are also featured on the
Online Books Page.}

The signal that a work---and, by extension, its author---is of interest to
contemporary audiences is only available for works in the public domain as
copyright monopolies limit the range of works that can be made into digital
editions.  For example, Project Gutenberg refuses works not in the public domain
in the United States and ebooks@Adelaide will only host works in the public
domain in Australia. (The Online Books Page lists digital editions of works
without regard for the legal jurisdiction of the hosting collection.) In order to apply this common standard
of notability more broadly, a strategy is needed to estimate the likelihood that
an author would have digital editions of their work(s) absent legal restrictions
on the dissemination of digital editions.  Combining data from the Online Books
Page with data from Wikipedia enables such an inference.

The content of Wikipedia articles and the record of reader and editorial
activity provide a rich source of data about individuals who have Wikipedia
pages devoted to them, including authors of literary, technical, and scientific
works. \numwikipediaindividuals{} individuals, authors and non-authors, have
a biographical Wikipedia article.\footnote{Unless otherwise noted, all
    references are to a Wikipedia ``dump'' made on April 2nd, 2014. To be
    included in the dataset, an individual's Wikipedia page must have one or
    more of the following: a birth date, a death date, or a bibliographic
    identifier. The bibliographic identifiers considered are BNF, GND, ISNI,
    LCCN, NLA, SELIBR, ULAN, and VIAF. In the interest of having data that are
modestly homogeneous, individuals who died before the year 1000 are not included
unless they have a bibliographic identifier. Data from the Online Books page was gathered on May 16th, 2014.} The range of data associated with
Wikipedia pages is considerable and while the body of active editors on
Wikipedia has shown deplorable biases---in a 2011 survey, only 9\% of Wikipedia
editors were women---many pieces of information, notably article age, provide
useful indicators of contemporary interest. For example, even if articles devoted to women are
systematically shorter, less frequently edited, and more likely to focus on
personal details, it might nevertheless be the case that women writers
in whom there is a strong contemporary interest will have pages which were
\emph{started} at an earlier date than pages about male writers in whom there is not as
strong an interest \citep{wikimedia_foundation2011editor, lam2011wpclubhouse,
    hill2013wikipedia, reagle2011gender, bamman2014unsupervised}.

The strategy pursued in this paper uses several streams of data associated with
individuals' Wikipedia pages to assess how strongly biographical articles resemble
articles concerning authors whose works have public domain digital editions. The data
used include the textual content of the article as well as the historical record
of reader and editorial activity linked with a page. For each individual's page,
the following features are extracted from a Wikipedia snapshot and page view
records: article length, article age in days, time elapsed since last revision,
revision rate during article's life, article text features (\numtopics topic weights
derived from a topic model), category count, translation count, redirect count,
estimated views per day, presence of translation for the \numtranslations
Wikipedias with the most translations, presence of bibliographic identifier
(GND, ISNI, LCCN, VIAF), article quality classification (``Good Article'' and
``Featured Article''), presence of protected classification, indicator for
decade of death for decades 1910--1950, and interactions between article age and
all features.\footnote{The number of page views per day is estimated from
a random sample of 60 days drawn between 2012-09-01 and 2013-08-31. Page views
of redirects are included.} The majority of the \numfeatures features used (not
counting interactions) are the \numtopics topic weights derived from
a non-parametric topic model \citep{buntine2014experiments}. The topic model is
fit with the entire corpus of \numwikipediaindividuals{} articles. Several topics
inferred from article texts are provided in Table~\ref{example-topics} along
with a sample of their characteristic words (stemmed).
Table~\ref{example-features} shows a subset of article features for several
familiar authors.

\begin{table}
\footnotesize
\begin{tabular}{ll}
\toprule
{} &                                                                             Characteristic words \\
\midrule
Topic &                                                                                                  \\
1     &  of the buddhist and swami buddhism spiritu burmes tibetan burma in tibet templ zen ethiopia sri \\
4     &     categori of birth death stub date name persondata place metadata peopl wikipedia defaultsort \\
31    &    painter paint of art artist the and in work museum portrait galleri exhibit sculptor sculptur \\
34    &    align center style text rowspan valign left bgcolor colspan width top right wikit td class br \\
35    &    he in his was and the to of categori at death die as from name after school famili son father \\
59    &    he the in of was and his categori to at birth death name for from as date place persondata on \\
64    &     danish denmark iceland copenhagen dk superliga hansen boldklub jen jensen dansk nielsen gibb \\
68    &    the of and librari in vol book societi publish org archaeolog collect archiv volum by histori \\
100   &    the book writer novel fiction of and stori isbn novelist author publish write in for award by \\
149   &            of the and in historian univers languag histori studi translat book scholar professor \\
168   &                     the to that in and of ref was had by his for not it as would on be with were \\
\bottomrule
\end{tabular}

\caption{Examples of topics derived from text of Wikipedia articles.}\label{example-topics}
\end{table}

\begin{table}
\footnotesize\begin{tabular}{lrrrrrrrr}
\toprule
                   &  Views &     \ding{61} &  \# redirect &  \# trans. &  Length (log) &   Age &  Rev./day &  Digital ed.? \\
\midrule
                   &        &       &             &           &               &       &           &               \\
 Charles de Gaulle &   1310 &  1970 &          26 &       114 &            12 &  4352 &       1.0 &             0 \\
      Christa Wolf &     38 &  2011 &           1 &        35 &             9 &  3827 &       0.1 &             0 \\
     George Orwell &   2481 &  1950 &          11 &       100 &            12 &  4618 &       1.7 &             1 \\
      Grace Hopper &    565 &  1992 &          12 &        42 &            10 &  4575 &       0.4 &             0 \\
     Hélène Cixous &      7 &   NA &           3 &        18 &            10 &  4098 &       0.1 &             0 \\
     J. K. Rowling &   4087 &   NA &          46 &        91 &            12 &  4575 &       1.6 &             0 \\
            Lu Xun &    186 &  1936 &          12 &        93 &            10 &  4265 &       0.2 &             1 \\
       Marie Curie &   3162 &  1934 &          31 &       134 &            11 &  4539 &       0.9 &             1 \\
      Ruth Rendell &    181 &   NA &           9 &        19 &            10 &  3862 &       0.1 &             0 \\
     Stieg Larsson &    811 &  2004 &           6 &        45 &            10 &  2514 &       0.3 &             0 \\
       T. S. Eliot &   1664 &  1965 &          25 &        81 &            11 &  4551 &       1.0 &             1 \\
       Thomas Mann &    573 &  1955 &           3 &        85 &            10 &  4694 &       0.2 &             1 \\
    Virginia Woolf &   1902 &  1941 &          14 &        76 &            11 &  4497 &       0.6 &             1 \\
\bottomrule
\end{tabular}

\caption{Subset of features derived from Wikipedia associated with familiar individuals.}\label{example-features}
\end{table}

Using the Online Books Page, information about which individuals' works have digital editions
is collected for all individuals who have Wikipedia pages and who died between
January 1st, 1910 and December 31st, 1952. Any work created by these individuals
entered the public domain before 2003 in 50 year p.m.a. countries such as
Australia, Canada, and Japan.\footnote{The cutoff of 2003 occurs before
Australia's modification of its copyright law in 2006. In that year the
country moved to a 70 year p.m.a. term. Any work entering the public domain
prior to 2003 according to a 50 year p.m.a. term is likely to be in the
public domain in Australia. Other notable countries with a 50 year p.m.a.
term include China, Hong Kong, Egypt, Iran, Indonesia, Japan, New Zealand,
South Africa, and South Korea.  Constraining the dataset to authors who died
after December 31st, 1909 also limits the heterogeneity of the authors
considered and the burden of double checking matches between Wikipedia
articles and records of digital editions on the Online Books Page.} In these countries more
than a decade has passed during which those interested in an author's works had
the opportunity to create digital editions. Inspecting a list of editions recently produced by Project
Gutenberg Canada, Project Gutenberg Australia, and ebooks@Adelaide reveals that this
opportunity has not been wasted. (For authors who died before 1928, more than
a decade has elapsed during which digital editions might have been made in 70
year p.m.a countries as well.) Table~\ref{obp-percentage} shows the percentage of
individuals with Wikipedia pages who have at least one digital edition listed in
the Online Books Page.

\begin{table}
\begin{tabular}{lr}
\toprule
{} &  Percentage having digital edition(s) \\
\midrule
1910-1919 &                                   3.2 \\
1920-1929 &                                   3.6 \\
1930-1939 &                                   3.2 \\
1940-1949 &                                   2.1 \\
1950-1952 &                                   2.3 \\
\bottomrule
\end{tabular}

\caption{Percentage of individuals on Wikipedia with indicated death years having at least one digital edition listed on The Online Books Page ($n$ = 85,424).}\label{obp-percentage}
\end{table}

\section{Model}\label{model}

Logistic regression is used to estimate the importance of a biographical article's
features in predicting the existence of one or more digital editions of an
author's works. To guard against over-fitting in the presence of a large number
of covariates, weakly informative prior distributions are given to the regression
coefficients \citep{gelman2008weakly}.\footnote{Following
\citet{gelman2008weakly}, explanatory variables such as those listed
are normalized to have mean zero and standard deviation one. The prior
distributions on the coefficients are Student-t distributions, $\beta \sim
t_7(0, 5)$.  This prior may also be understood as a form of regularization.
Computation of posterior distributions, unless otherwise noted, uses Hamiltonian Monte Carlo
\citep{stan_development_team2014stan}.} After fitting the model with data from
the \numwikipediaindividualstrain individuals who died between 1910 and 1953,
the fitted regression coefficients are used, along with the entire dataset
covering \numwikipediaindividuals{} individuals, to calculate the posterior
distribution of the predictive probability of an individual having at least one
digital edition (assuming, for the sake of the exercise, a world without
obstacles to the creation of digital editions). An individual's
Public Domain Rank is an individual's expected rank in terms of these predicted
probabilities. An alternative measure, the Public Domain Score, is a number
between 0 and 100 corresponding to the expected quantile (multiplied by 100) of
the predicted probability. For example, the author Virginia Woolf has a Public
Domain Rank of \woolfrank{} (out of \numwikipediaindividuals{}) and a Public Domain
Score of \woolfscore{}. By contrast, Giuseppe Amisani, an Italian painter who died
in the same year as Woolf, has a rank of \amisanirank{} and score of
\amisaniscore{}.

\begin{figure}
  \centering
    \includegraphics[width=0.9\textwidth]{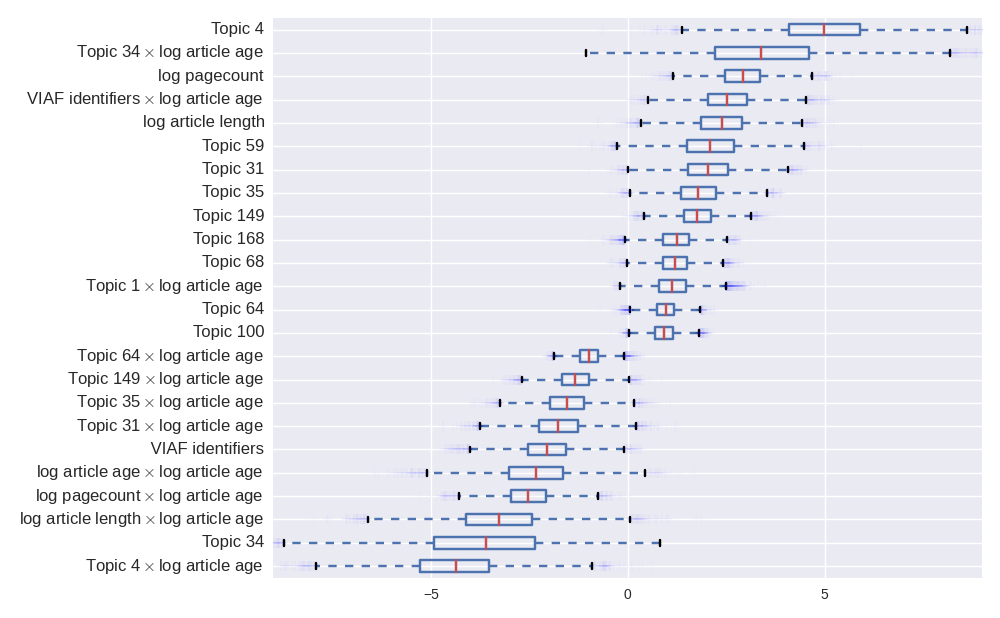}
    \caption{Coefficients of the regression of the presence of a digital edition
    on selected features.}\label{coefs}
\end{figure}

Although a thorough understanding of the relationship between Wikipedia
editorial activity and the curatorial activity surrounding the cultural commons
is beyond the scope of this paper, inspecting the posterior distributions of the
regression coefficients yields modest insights (Figure~\ref{coefs}). For
example, while well-known painters will reliably have comprehensive Wikipedia
articles, they do not have public domain editions of their works according to
the definition used in this paper. This state of affairs is reflected by the
large negative coefficient associated with the interaction between article age
and topic 31 (``painter'', ``paint'', ``art''). The positive predictive value of
topic 100 comes as no surprise, as the words associated with the topic include:
``book'', ``writer'', ``novel'', and ``fiction.'' The presence of topic 4
appears to indicate a well-tended biographical article; ``persondata'' and
``defaultsort'' are template names commonly used in biographical articles.

A qualitative sense of the performance of the model may be gained by examining
Table~\ref{notable-2015-2025}, which shows the top ranked authors whose works
will enter the public domain between 2015 and 2025 in 50 year p.m.a. jurisdictions.
(The ranking of all individuals is available at \url{\websiteurl}.)

\section{Evaluation}\label{evaluation}

Cross-validation provides a basic check of the reliability of the model. In
section~\ref{comparison}, the results of the model are compared with existing
rankings of authors and literary works.

\begin{figure}[htbp]
\centering
\includegraphics[height=4in]{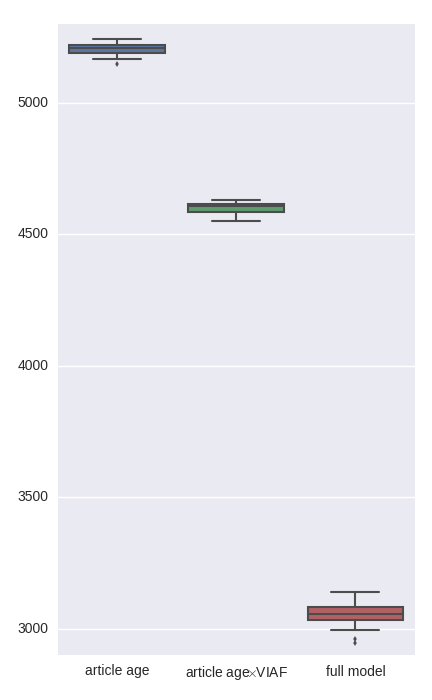}
\caption{Comparison of full model with baseline models using log loss. A lower
score indicates better out-of-sample predictions.}\label{cross-validation}
\end{figure}

Cross-validation is performed in the following way. A model is given data for
a subset of the \numwikipediaindividualstrain individuals who died between 1910
and 1953. By turns, information concerning one half of these individuals,
including whether or not there are digital editions associated with them, is
``held out'' and the model is fit with the remaining data. The model then makes
predictions for the held-out portion and the accuracy of these predictions is
assessed using a loss function, the log loss.\footnote{The log loss of
a prediction is $\mathcal{L}(y, \hat p) = - \sum_{n=1}^N y_n \times \log
\hat p_n + (1 - y_n) \times \log \hat p_n$, where $y_n$ is a zero or a one
indicating whether a digital edition associated with author $n$ exists and
$p_n$ is the predicted probability that such a work exists. To avoid the
computational costs associated with calculating the expected log loss,
predictions for cross-validation are calculated using a point estimate for
all models.} This process is repeated twenty times with a different half
being chosen to hold out each time. The baseline models to which the full
model are compared include the following: a model using only article age and
a model using article age, the presence of a VIAF bibliographic identifier
(common for authors), and the interaction of the two features.
Figure~\ref{cross-validation} shows the results of the cross-validation and
confirms that the full model makes better predictions than the baseline models.

\section{Comparison with Existing Rankings}\label{comparison}

In 1998 the Modern Library's editorial board collected a list of the 20th
century's ``best'' works of fiction
\citep{modern_library_editorial_board1998100}. At the time, the editorial board
featured luminaries such as Gore Vidal and A. S. Byatt.  The Public Domain Rank
of the authors of works selected by the Modern Library's editorial board are consistently
high (Table~\ref{modernlibrary}). The median rank of the authors whose works
were selected is \medianmodernlibrary (of \numwikipediaindividuals). The only
outlier is Samuel Butler (1835–-1902), who was selected for the posthumously
published \emph{The Way of All Flesh} (1903).

\begin{table}
\footnotesize\begin{tabular}{lr}
\toprule
                          & Public Domain Rank (Score) \\
\midrule
          Ford Madox Ford &                   813 (94) \\
            Joseph Conrad &                  1358 (94) \\
            Edith Wharton &                  2098 (94) \\
              James Joyce &                  2799 (94) \\
             Wilra Cather &                  3164 (94) \\
 Wilriam Kennedy (author) &                  3179 (94) \\
              Saul Belrow &                  3474 (94) \\
            Robert Graves &                  3483 (94) \\
         Theodore Dreiser &                  4037 (94) \\
      W. Somerset Maugham &                  4107 (94) \\
             Jack Kerouac &                  4186 (94) \\
             Henry Milrer &                  4819 (93) \\
             Walker Percy &                  6864 (93) \\
          Anthony Burgess &                  8035 (93) \\
          John Dos Passos &                  9326 (93) \\
          Arthur Koestler &                  9332 (93) \\
            Joseph Helrer &                 18331 (91) \\
                Jean Rhys &                 33258 (89) \\
         Lawrence Durrelr &                 42740 (88) \\
 Samuel Butler (novelist) &                304859 (58) \\
\bottomrule
\end{tabular}

\caption{Public Domain Rank for a random sample of authors whose works were selected by the Modern Library's editorial board for inclusion in a list of the 20th century's best works of fiction.}\label{modernlibrary}
\end{table}

A second validation of the ranking draws on Canadian legal scholar Michael
Geist's list of notable Canadian authors whose works will, absent change in
current law, enter the public domain between 2014 and 2020 \citep{geist2012tpp}
(Table~\ref{canadian-authors-geist}).  Apart from the low ranking of the
classical musician and novelist Winifred Bambrick, the ranking aligns favorably
with Geist's list. The ranking also provides a way of finding prominent
Canadians Geist omits.  Well-known Canadian authors not appearing on the list
include Chief Dan George, John Russell Harper, Yves Thériault, and
Félix-Antoine Savard.\footnote{In this case, a list of individuals with the
category ``Officers of the Order of Canada'' was sorted by Public Domain Rank
and inspected for authors not on Geist's list.}

\begin{table}
\footnotesize\begin{tabular}{lrrr}
\toprule
                              &     \ding{61} &  Views & Public Domain Rank (Score) \\
\midrule
                Gabrielle Roy &  1983 &     38 &                  3133 (94) \\
             Donald Creighton &  1979 &      9 &                 10773 (92) \\
             Marshall McLuhan &  1980 &    629 &                  2299 (94) \\
             Gwethalyn Graham &  1965 &      2 &                 39931 (89) \\
                 Hubert Aquin &  1977 &      9 &                 65185 (85) \\
                 Ethel Wilson &  1980 &      6 &                  8028 (93) \\
                  E. J. Pratt &  1964 &     24 &                  6074 (93) \\
 Susan Wood (science fiction) &  1980 &      5 &                  2005 (94) \\
            Winifred Bambrick &  1969 &      1 &                288058 (63) \\
        Winthrop Pickard Bell &  1965 &      2 &                 45017 (88) \\
            Thomas B. Costain &  1965 &     24 &                  3993 (94) \\
     Ralph Allen (journalist) &  1966 &      1 &                  6107 (93) \\
                  Hugh Garner &  1979 &      9 &                  8268 (93) \\
          Germaine Guèvremont &  1968 &      0 &                 23793 (91) \\
                  A. M. Klein &  1972 &     16 &                  8357 (93) \\
\bottomrule
\end{tabular}

\caption{Notable Canadian authors whose works will likely enter the public domain in Canada between 2014 and 2020.}\label{canadian-authors-geist}
\end{table}

In general, Public Domain Rank reflects received judgements. Allowing for its
tendency to mix the sacred and the profane (from the perspective of the Modern
Library's editorial board) it tends to rank familiar figures in the popular and
literary firmament very high. With rare exceptions, Public Domain Rank also
proves able to identify notable writers who did not write in English
(Table~\ref{rankings-nonamerican-nonbritish}).

\begin{table}\footnotesize
\begin{tabular}{lrr}
\toprule
                                &  Views & Public Domain Rank (Score) \\
\midrule
                                &        &                            \\
                    Anita Desai &    185 &                  5731 (93) \\
                   Tove Jansson &    193 &                  6756 (93) \\
                    Thomas Mann &    573 &                  7418 (93) \\
                  Edogawa Ranpo &     58 &                  8376 (93) \\
                        Lao She &     40 &                 16954 (92) \\
       Antoine de Saint-Exupéry &     19 &                 18168 (91) \\
                         Lu Xun &    186 &                 25148 (90) \\
                  Alfred Döblin &      3 &                 28681 (90) \\
                    Agnar Mykle &      5 &                 28776 (90) \\
                       Raja Rao &     49 &                 30326 (90) \\
                  R. K. Narayan &    547 &                 40650 (88) \\
                   Christa Wolf &     38 &                 43706 (88) \\
                    Octavio Paz &    231 &                 46852 (88) \\
                 Carlos Fuentes &    223 &                 48992 (87) \\
                 Mulk Raj Anand &    103 &                 77413 (84) \\
                 A. C. Baantjer &      6 &                106790 (80) \\
                Khushwant Singh &    310 &                112581 (80) \\
 Machado de Assis &     75 &                401021 (55) \\
\bottomrule
\end{tabular}

\caption{Rankings of selected authors who did not write in English or who published primarily outside the United States and United Kingdom.}\label{rankings-nonamerican-nonbritish}
\end{table}

\section{Public Domain Rank as a general index of notability}\label{analysis}\label{public-domain-rank-as-a-general-index}

Public Domain Rank also provides a general ranking of individuals. To the
extent that a biographical Wikipedia article (concerning a non-author) shares characteristics with
pages about individuals who have written works that have digital editions, the
individual will be ranked higher. This is not a controversial idea in itself;
it is not surprising that pages that, for example, have been around for longer
(article age) or which have been edited more recently would be of greater
contemporary interest. Public Domain Rank, however, provides a transparent and
reproducible method for assigning precise weights to such features. That the
ranking is independent of any specific set of individuals (such as an editorial
board or prize jury) should be weighed against the ranking's idiosyncrasies.
Consider, for example, the ranking of 20th-century British Prime Ministers
(Table~\ref{rankings-british-prime-ministers}). The ranking provided by Public
Domain Rank compares favorable with a ranking assembled by the University of
Leeds and Ipsos Mori in 2004, which was based on responses from 139 academics
who specialized in 20th-century British history and/or politics (respondents
were asked to judge the ``success'' of the politicians). Using a pairwise loss
function and the Leeds/Mori ranking as the standard, Public Domain Rank
performs better than a random ranking with 97\% confidence.
Indeed, it performs better than a ranking of the Prime Ministers in terms of
page views.

\begin{table}\footnotesize
\begin{tabular}{lrr}
\toprule
                                                  &  Leeds 2004 & Public Domain Rank (Score) \\
\midrule
                                                  &             &                            \\
                                Winston Churchill &           2 &                  4557 (93) \\
                                Margaret Thatcher &           4 &                  5342 (93) \\
                                   Clement Attlee &           1 &                  9901 (93) \\
                                 Ramsay MacDonald &          14 &                 11077 (92) \\
                                       Tony Blair &           6 &                 12774 (92) \\
                                   Arthur Balfour &          18 &                 14155 (92) \\
                                     Edward Heath &          13 &                 14202 (92) \\
                                     Anthony Eden &          20 &                 14439 (92) \\
                                 Harold Macmillan &           5 &                 14581 (92) \\
                                    H. H. Asquith &           7 &                 15128 (92) \\
                               David Lloyd George &           3 &                 15515 (92) \\
                                    Harold Wilson &           9 &                 17180 (91) \\
                                  James Callaghan &          12 &                 18130 (91) \\
                                  Stanley Baldwin &           8 &                 18952 (91) \\
                              Neville Chamberlain &          17 &                 22332 (91) \\
                                Alec Douglas-Home &          19 &                 26231 (90) \\
                                       John Major &          15 &                 27296 (90) \\
                                        Bonar Law &          16 &                 28735 (90) \\
 Robert Gascoyne-Cecil &          10 &                272898 (59) \\
                         Henry Campbell-Bannerman &          11 &                290128 (58) \\
\bottomrule
\end{tabular}

\caption{Ranking of 20th-century British Prime Ministers.}\label{rankings-british-prime-ministers}
\end{table}

\section{Public Domain Rank's Biases}\label{public-domain-ranks-biases}

Public Domain Rank faithfully reflects the biases (or, the wisdom) of
the relevant ``crowd'': volunteers creating digital editions. These biases are
not difficult to identify. A casual inspection of the many texts on, for
example, Project Gutenberg reveals that, alongside well-known popular and
canonical writers, many works have a political or religious character. Works
belonging to popular genres such as mystery, science fiction, and fantasy also
appear frequently. Public Domain Rank reproduces these biases. Indeed, perhaps
the surest route to the higher reaches of Public Domain Rank is to be a writer
of popular fiction addressing political or religious issues.

The weaknesses of Public Domain Rank come from two sources, (1) the
particularities of the input to the model (the population of digital editions)
and (2) bias in the coverage of individuals in the English-language Wikipedia.
Both of these shortcomings will likely be ameliorated over time. The English-language
Wikipedia continues to expand, notably via the route of having significant
articles that appear in other languages' Wikipedias translated into English.
Even a limited exposure to Wikipedia should persuade one that it is likely that
demonstrably famous individuals, regardless of their country of origin, will
tend to find their way into the English-language Wikipedia.  Many well-known
authors who never wrote in English already have substantial pages on the
English-language Wikipedia. With regards to the ``input,'' the set of digital
editions, it seems likely that in ten years the range of authors who
died between 1910 and 1953 who have digital editions will have expanded
considerably. Were the model updated with Wikipedia and Online Books Page
snapshots in ten years time, Public Domain Rank would likely better capture the
contemporary importance of individuals who did not write in English.

The essential input to Public Domain Rank are the individual decisions of
volunteers. These volunteers elect to participate in the creation of digital
editions on sites such as Distributed Proofreaders. Volunteers in a position to make
such a contribution currently come from a biased sample of the global population
of readers. For example, the organizations and collectives most successful at
attracting and organizing the efforts of volunteers---such as Project Gutenberg
and Distributed Proofreaders---have tended to produce English-language texts. As other
countries expand digitization efforts and internet access, works of enduring
popularity in languages other than English will become available in digital
editions and will find their way onto the Online Books Page. Neither Project Gutenberg nor the
Online Books Page is restricted to English-language works. For example, The Online
Books Page has entries for the following digital editions of the works of Lu Xun
(1881-1936) (all Gutenberg editions): 吶喊, 中國小說史略, 狂人日記, and
南腔北調集.

It is possible, however, that the biases may grow worse. For example, the
Mormon Church, already a significant source of public domain digital editions
of works relevant to its members, might decide to subsidize the digitization of
all public domain works by Mormon authors. The
\href{http://www.sfwa.org/}{Science Fiction \& Fantasy Writers of America}
might successfully brigade fans of American science fiction into creating
digital editions of all works of science fiction published in the United States
before 1940. While these developments would be welcome, they would call into
question the foundations of Public Domain Rank.  If such developments were to
occur, however, one response would be to explicitly model the ways in which
those contributing public domain editions fail to represent the global
population of readers. Collecting demographic information about a random sample
of contributors of digital editions would be neither time-consuming nor
expensive.

\section{Applications}
\subsection{Automating Public Domain Day}\label{automating-public-domain-day}

Public Domain Rank promises to facilitate---and even automate---Public Domain
Day. Table~\ref{notable-2015-2025} shows the top ranked authors whose works
will enter the public domain between 2015 and 2020. Such a list may be generated
for an arbitrary year. For example, a group interested in compiling a list for
Public Domain Day 2020 in Europe will be interested in identifying notable
authors whose works enter the public domain in 2020 in 70 year p.m.a. countries.
By consulting the ranking of individuals who died in 1949 the group can ensure they
have not overlooked any obvious candidates.

\begin{table}
\footnotesize\begin{tabular}{lrrr}
\toprule
                         &     \ding{61} &  Views & Public Domain Rank (Score) \\
\midrule
                         &       &        &                            \\
 Martin Luther King, Jr. &  1968 &   9160 &                   119 (95) \\
          August Derleth &  1971 &     77 &                   342 (95) \\
          Margaret Irwin &  1969 &      6 &                   430 (94) \\
           Fredric Brown &  1972 &     69 &                   442 (94) \\
           Bruce Elliott &  1973 &      3 &                   456 (94) \\
           Groff Conklin &  1968 &      6 &                   459 (94) \\
      Robert Arthur, Jr. &  1969 &     23 &                   534 (94) \\
         Anthony Boucher &  1968 &     23 &                   602 (94) \\
       Elizabeth Enright &  1968 &     13 &                   715 (94) \\
          Conrad Richter &  1968 &     20 &                   732 (94) \\
        J. R. R. Tolkien &  1973 &   4634 &                   733 (94) \\
      Rosel George Brown &  1967 &      2 &                   773 (94) \\
     Charlotte Armstrong &  1969 &     13 &                   995 (94) \\
             T. S. Eliot &  1965 &   1664 &                  1094 (94) \\
        John W. Campbell &  1971 &     80 &                  1099 (94) \\
       Margery Allingham &  1966 &     49 &                  1164 (94) \\
        Charles Beaumont &  1967 &     38 &                  1268 (94) \\
       Flannery O'Connor &  1964 &     15 &                  1279 (94) \\
             Ruth Sawyer &  1970 &      6 &                  1281 (94) \\
             Enid Blyton &  1968 &    746 &                  1362 (94) \\
         David H. Keller &  1966 &      9 &                  1397 (94) \\
            Allan Seager &  1968 &      3 &                  1551 (94) \\
            Harl Vincent &  1968 &      2 &                  1581 (94) \\
         Shirley Jackson &  1965 &    244 &                  1605 (94) \\
          Upton Sinclair &  1968 &    594 &                  1616 (94) \\
             W. H. Auden &  1973 &    615 &                  1638 (94) \\
         Eleanor Farjeon &  1965 &     42 &                  1673 (94) \\
        Vincent Starrett &  1974 &      6 &                  1784 (94) \\
            Philip Wylie &  1971 &     28 &                  1800 (94) \\
             T. H. White &  1964 &     90 &                  1834 (94) \\
\bottomrule
\end{tabular}

\caption{Notable authors with works entering the public domain between 2015 and 2025 in 50 year p.m.a. copyright-term countries.}\label{notable-2015-2025}
\end{table}

\subsection{Expanding the Commons}\label{expanding-the-commons}

Users of this ranking include those those organizing efforts to digitize works
that are in the public domain but lack digital editions. Public Domain Rank
facilitates identifying authors whose works, were they made
available in digital editions, would likely find an audience.
Table~\ref{public-domain-opportunities} shows authors whose works are in the public
domain in 50 year p.m.a. countries but lack public domain digital
editions. Flannery
O'Connor and Sylvia Plath stand out as significant examples of authors whose
works might be made available today on Project Gutenberg Canada. Richard Wright
(author of \emph{Native Son}) also ranks highly.

\begin{table}
\footnotesize\begin{tabular}{lrrr}
\toprule
                     &     \ding{61} &  Views & Public Domain Rank (Score) \\
\midrule
                     &       &        &                            \\
 Otis Adelbert Kline &  1946 &      9 &                   618 (94) \\
  John Russell Fearn &  1960 &      4 &                   749 (94) \\
    Dashiell Hammett &  1961 &    308 &                   763 (94) \\
  Henry S. Whitehead &  1932 &      4 &                   826 (94) \\
        Mark Clifton &  1963 &      4 &                  1111 (94) \\
 Margaret Wise Brown &  1952 &     52 &                  1147 (94) \\
    Arthur Leo Zagat &  1949 &      3 &                  1181 (94) \\
   Flannery O'Connor &  1964 &     15 &                  1279 (94) \\
      Fletcher Pratt &  1956 &     12 &                  1310 (94) \\
         Tod Robbins &  1949 &      8 &                  1311 (94) \\
    Robert E. Howard &  1936 &    306 &                  1335 (94) \\
      Gerald Bullett &  1958 &      3 &                  1347 (94) \\
     Miles J. Breuer &  1945 &      4 &                  1367 (94) \\
 Julian Osgood Field &  1925 &      2 &                  1609 (94) \\
    Lavinia R. Davis &  1961 &      3 &                  1655 (94) \\
    Coulson Kernahan &  1943 &      1 &                  1671 (94) \\
       Sarah Doudney &  1926 &      2 &                  1747 (94) \\
     Oscar J. Friend &  1963 &      2 &                  1811 (94) \\
         T. E. Hulme &  1917 &     28 &                  1833 (94) \\
        Rachel Field &  1942 &     16 &                  1901 (94) \\
       D. K. Broster &  1950 &      5 &                  1980 (94) \\
      James Gairdner &  1912 &      3 &                  2372 (94) \\
        Sylvia Plath &  1963 &   2101 &                  2394 (94) \\
\bottomrule
\end{tabular}

\caption{Authors whose works are in the public domain in 50 year p.m.a. countries but whose works lack digital editions.}\label{public-domain-opportunities}
\end{table}

\subsection{Overlooked Wikipedia articles}

Public Domain Rank serves to support improving Wikipedia by
identifying biographical articles deserving attention.  Discerning where Wikipedia editors should
spend their available time is a task that Wikipedia editors have identified as
important. Numerous WikiProjects (groups of contributors working as a team) rate
articles in terms of their ``importance'', assigning articles a rating (``Low'',
``Mid'', ``High'', and ``Top'').  Public Domain Rank provides an independent
rating of an individual's importance to a contemporary audience and may support
Wikipedia contributors in prioritizing their efforts.  For example, searching
for authors with high Public Domain Rank but whose articles were lacking
bibliographic identifiers revealed that the Wikipedia page associated with the
American novelist and essayist James Baldwin (1924-1987) lacked any form of
bibliographic identifier---something atypical for a writer of his stature.

\section{The Aesthetics of Public Domain Rank}\label{the-aesthetics-of-public-domain-rank}

Many of the individuals featured in the ranking are authors of works of poetry
and prose fiction. A majority of the public domain digital editions used as
input to the ranking are literary works. The ranking of (authors of) literary
works naturally raises the question of aesthetic judgement. What kind of judgement
is implied by Public Domain Rank? Is it a judgement deserving of our attention
as an aesthetic judgement? To a first approximation the ranking provided by Public Domain
Rank looks questionable. It is as if someone proposed awarding the Booker Prize based on the advice of
a committee of randomly selected book enthusiasts (e.g., those likely to contribute
their time to Project Gutenberg).

The aesthetic theory of David Hume (1711-1776, 8 digital editions listed in the
Online Books Page) is particularly relevant here. Hume, in contrast to his
contemporaries, draws our attention specifically to works with enduring
popularity. It is, of course, these works that Project Gutenberg and similar
sites aim to collect.

Hume's theory of taste is distinctive in that it does not argue for a
special faculty of taste or general principles characterizing the
beautiful (c.f., Immanuel Kant, Francis Hutcheson)
\citep{dickie1996century}. According to Hume, judgements of beauty or
ugliness are sentiments, \emph{tout court}
\citep[136]{hume2008selected}. But Hume rejects skepticism about taste. That is, he has something to
say to someone who wants to claim, after one distracted viewing, that
the latest made-for-television movie is aesthetically superior to Orson
Wells' \href{https://en.wikipedia.org/wiki/Citizen_Kane}{\emph{Citizen Kane}}
or Abbas Kiarostami's
\href{https://en.wikipedia.org/wiki/Taste_of_Cherry}{\emph{A Taste of Cherry}}.
Hume's principle argument here is that we frequently observe agreement
among critics about the beauty-making and beauty-destroying
characteristics of individual works of art as well as agreement about
specific (extreme) comparisons between two works of art (e.g., a novel
by Toni Morrison and a poorly written mystery novel).\footnote{The
  language ``beauty-making'' and ``beauty-destroying'' is borrowed from
  \citet{dickie1996century}.}

Hume holds up aesthetic judgements of beauty that are dispassionate,
attentive, and experienced \citep[139]{hume2008selected}. Judgements of
a work as beautiful or flawed which are made without bias are given
priority, as are those made by critics who have extensive experience
with the relevant kind of cultural artifact. An example of an assessment
of a literary work likely to deserve our attention would be, on this
account, that of a reader who had read widely and has no personal or
financial connection to the work or its creator.

Appealing to these considerations, Hume argues that cultural works that have
stood the test of time deserve credence because judgements of their aesthetic
characteristics are more likely to be dispassionate than judgements of more
recent works, as the latter are more likely to be compromised by contemporary
concerns such as personal connection or even jealousy \citep[139--40]{hume2008selected}.  To the extent that one finds critics equipped with extensive experience
with the relevant kind of cultural work, a ``perfect serenity of mind'',
and ``due attention to the object'' \citep[139, ¶10]{hume2008selected}
then one finds in their ``joint verdict'' the ``true standard of taste
and beauty'' \citep[147, ¶23]{hume2008selected}.

Superficially the ``joint verdict'' on the quality of a literary work represented by
its presence on Project Gutenberg appears to share many characteristics with the
standard of taste favored by Hume. The
requirement that the work be published more than 50 years ago, in particular,
supports dispassionate judgement insofar as judgements of recently published works
are more likely to be poorly considered. And
while there is no requirement that those proposing the creation of a
digital edition on a platform such as Distributed Proofreaders (the
principle contributor to Project Gutenberg) have read widely, one is
assured that the work has been given at least one person's
undivided attention in the process of manually keying-in the text. It seems
likely that a volunteer might abandon---or at least slacken---their efforts
if they came to the conclusion that the work occupying their time was,
on balance, more blemished than beautiful.\footnote{This point seems
  less persuasive with shorter works such as poetry, children's literature, and
  short stories. While typing in the contents of Tolstoy's \emph{War and
  Peace} is a task not be undertaken likely, a collection of poems might be
  entered and proofread in a few hours.} And while anyone may
propose and contribute a digitization to Project Gutenberg via
Distributed Proofreaders, the risk of books being selected haphazardly,
without ``due attention to the object'' is limited by
institutional barriers \citep[139]{hume2008selected}: in order to
shepherd works through the digitization process at Distributed
Proofreaders, a user must pass a number of formal quizzes and contribute
a considerable amount of work to existing projects.\footnote{\href{http://www.pgdp.net/wiki/Getting_started}{``Getting
  Started - DPWiki.''} Accessed August 15, 2014.
  http://www.pgdp.net/wiki/Getting\_started.}

This is not to say that bias of the kind that concerned Hume is absent from
the creation of digital editions and therefore absent in Public Domain Rank.
Even if we accept that the ranking reflects which works command more than
a casual interest among contemporary readers, the interest may reflect personal
connections with the author or work.  The volunteer digitizing the work may have
known the author in question.  For example, we have witnessed the digitization
of academic articles of a parent by a child.\footnote{``The Tree of Life:
Freeing My Father's Publications.'' Accessed August 21, 2014.
http://phylogenomics.blogspot.co.uk/p/freeing-my-fathers-publications.html.} In
other cases, the volunteers may be connected with an organization founded by the
original author. Religious organizations coordinating the creation of digital
editions provide a prominent example of this.\footnote{For example, a variety of
works associated with the Mormon Church and the Hare Krishna Movement have public
domain digital editions listed on the Online Books Page.}

\section{Conclusion}\label{conclusion}

This paper introduces and validates an automatic method for identifying notable
individuals, where notability is defined using records public domain digital
editions. This bottom-up approach to identifying works and individuals of
enduring interest makes use of two sources of open data, the Online Books Page
and Wikipedia. By aligning bibliographic records in the Online Books Page with
the streams of structured and unstructured data from Wikipedia, this project
facilitates the identification of notable works in the public domain.

\section*{Acknowledgements}\label{acknowledgements}

I would like to acknowledge the computational resources provided by the Neukom
Institute and the Math Department of Dartmouth College and the assistance
provided by Sarunas Burdulis. John Mark Ockerbloom, Sravana Reddy, Kirstyn
Leuner, and Kes Schroer provided valuable feedback on drafts of this paper.

\end{CJK}

\bibliographystyle{plainnat}
\bibliography{bibliography}
\end{document}